\documentclass[prb,twocolumn,showpacs,amsmath,amssymb,superscriptaddress]{revtex4-2}
\usepackage{amsmath,amssymb}
\usepackage[pdftex]{hyperref,graphicx}
\hypersetup{colorlinks = true, urlcolor = blue, linkcolor = blue, citecolor = blue}
\usepackage{physics}
\usepackage{xcolor}
\usepackage{bm}

\newcommand{\comment}[1]{{}}

\newcommand{\cmmnt}[1]{}

\begin{document}
\title{Density-tuned effective metal-insulator transitions in 2D semiconductor layers: Anderson localization or Wigner crystallization?}

\author{Seongjin Ahn}
\affiliation{Condensed Matter Theory Center and Joint Quantum Institute, Department of Physics, University of Maryland, College Park, Maryland 20742-4111, USA}
\author{Sankar Das Sarma}
\affiliation{Condensed Matter Theory Center and Joint Quantum Institute, Department of Physics, University of Maryland, College Park, Maryland 20742-4111, USA}

\date{\today}

\begin{abstract}
Electrons (or holes) confined in two-dimensional (2D) semiconductor layers have served as model systems for studying disorder and interaction effects for almost 50 years.  In particular, strong disorder drives the metallic 2D carriers into a strongly localized Anderson insulator (AI) at low densities whereas pristine 2D electrons in the presence of no (or little) disorder should solidify into a Wigner crystal (WC) at very low carrier densities in order to optimize their Coulomb potential energy.  The ability to tune the carrier density continuously in a fixed sample allows the 2D semiconductor system to go from a high-density metallic Fermi liquid to a low-density disorder dominated Anderson insulator, or, if the sample is particularly clean, to a Coulomb interaction dominated low-density quantum Wigner crystal.  Since the disorder in 2D semiconductors is mostly Coulomb disorder arising from random unintentional quenched charged impurities in the environment, the applicable physics is complex as the carriers interact with each other as well as with the random charged impurities through the same long-range Coulomb coupling.  In addition, the Wigner crystallization occurs at such low carrier densities, that in most situations the relevant carrier density is comparable to the background charged impurity density even in ultraclean samples. By critically theoretically analyzing the experimental transport data in depth using a realistic transport theory to calculate the low-temperature 2D resistivity as a function of carrier density in 11 different experimental samples covering 9 different materials, we establish, utilizing the Ioffe-Regel-Mott (IRM) criterion for strong localization, a direct connection between the critical localization density for the 2D metal-insulator transition (MIT) and the sample mobility deep into the metallic state, which for particularly clean samples could lead to a localization density low enough to make the transition appear to be a Wigner crystallization.  We believe that the insulating phase is always an effective Coulomb disorder-induced strongly localized AI, which may have short-range WC-like correlations at very low carrier densities.  Our theoretically calculated disorder-driven critical MIT density agrees well with experimental findings in all 2D samples, even for the ultra-clean samples where the critical density approaches the WC transition density.  In particular, the extrapolated critical density for the 2D MIT seems to vanish when the high-density mobility goes to infinity, indicating that transport probes a disorder-localized insulating ground state independent of how low the carrier density might be.  

\end{abstract}

\maketitle
\section{Introduction} \label{sec:1}
Phases (e.g., crystals, fluids, ferromagnets, paramagnets, superconductors, insulators, .....) and transitions among them are the central themes of condensed matter physics.  Typically, phase transitions are thermodynamically driven by temperature at a critical temperature ($T_c$), but quantum phase transitions at $T=0$ are also possible, tuned by system parameters such as density, doping, electric field, magnetic field, pressure, etc.  An early quantum phase transition prediction by Wigner is that an electron Fermi liquid (FL) interacting via the long range Coulomb interaction crystallizes into a quantum `Wigner crystal'(WC) at a sufficiently low electron density ($n$) at $T=0$ \cite{wignerInteractionElectronsMetals1934}. Electrons in normal metals remain a Fermi liquid (assuming no superconductivity) even at $T=0$ because of the strong zero point energy associated with the finite Fermi energy \cite{pinesTheoryQuantumLiquids2018}. But with decreasing $n$, the relative energy cost (i.e. the ratio of the Coulomb potential energy to the kinetic energy $E_\mathrm{F}$) increases in an electron liquid as $r_s$, where $r_s \sim n^{-1/d}$ is the dimensionless Wigner-Seitz radius (with $n$ the electron density in $d$ dimensions) defined as the average electron separation measured in the effective Bohr radius. 
The precise definition of $r_s$ for our 2D systems of interest is $r_s=(\pi n)^{-1/2}/(\kappa\hbar^2/me^2)$, where kappa and m are respectively the background lattice dielectric constant and the carrier effective mass. Typical $r_s$ values of interest for 2D MIT being discussed in the current work depend strongly on the effective amount of disorder in the system and varies between $r_s=4$ and 50 for the systems studied in the current work, wit higher (lower) $r_s$ applying to lower (higher) disorder in the system (see Table 1 for the details).
Therefore, at some low (high) $n$ ($r_s$), the electrons should solidify into a WC to minimize its potential energy at some cost to the kinetic energy. In 2D, the critical $r_s$ value ($r_c$) for the FL-WC transition in clean systems at $T=0$ has been calculated by many groups through quantum Monte Carlo (QMC) simulations, and although different simulations obtain somewhat different results, $r_c \sim 30-40$ is accepted to be a reasonable theoretical estimate \cite{tanatarGroundStateTwodimensional1989, rapisardaDiffusionMonteCarlo1996, falakshahiHybridPhaseQuantum2005, drummondPhaseDiagramLowDensity2009}.
The same liquid-crystal transition, of course, would also happen thermodynamically with the creation of a classical WC with the lowering of temperature at a fixed low density where the electrons are nondegenerate.  Such a classical WC was indeed experimentally observed a long time ago for two-dimensional (2D) nondegenerate electrons confined on the surface of liquid He$_4$ by the image force \cite{grimesEvidenceLiquidtocrystalPhase1979}.
We do not consider the thermodynamic finite-$T$ transition at all in the current work.
Our interest in the current work is a theoretical understanding of density-tuned low-$T$ experimental transport properties of 2D carriers localized in various semiconductor layers \cite{andoElectronicPropertiesTwodimensional1982, dassarmaElectronicTransportTwodimensional2011a}, 
which are occasionally claimed to reflect an underlying WC transition by virtue of the extreme low density where the system manifests an effective metal-insulator transition from a higher-density  metallic transport behavior to a lower density insulating  transport behavior. Since dc transport can only distinguish between conducting (finite conductivity at $T=0$) and insulating (zero conductivity at $T=0$) phases, our work focuses on a critical analyses of a large set of experimental 2D transport publications reporting density-tuned effective metal-to-insulator transition (2D MIT), concluding that such generically observed density-tuned 2D MIT is more appropriately described as a disorder-induced Anderson (conductor-to-insulator) localization crossover rather than a correlation-induced (electron liquid-to-electron solid) WC transition.  

Of course, both disorder and correlation effects are simultaneously present in any sample, but the important question is whether a particular low-temperature density-tuned 2D MIT experimental observation should be construed as the manifestation of disorder-induced Anderson localization or interaction-induced Wigner crystallization since the data only reflect a sharp density-tuned crossover from a higher density ($n>n_c$) metallic resistivity (i.e. resistivity being $T$-independent at low enough $T$) to a lower-density insulating resistivity behavior (i.e. resistivity decreasing exponentially with increasing $T$) around a critical density $n_c$. The observed experimental behavior, although sharp, happening within a narrow density range of $n_c$, is consistent with a crossover (rather than a phase transition) with the resistivity around $n_c$ manifesting a complicated sample-dependent non-universal temperature dependence  which cannot clearly be characterized as conducting or insulating. However, the resistivity behavior for $n\gg  (\ll ) n_c$ is unambiguously metallic (insulating) at low temperatures, thus allowing for an estimation of the crossover critical density $n_c$.

By using the Boltzmann transport theory taking into account carrier scattering by screened random charged impurity scattering, we calculate the system resistivity in the metallic phase ($n>n_c$) as a function of carrier density, and then obtain the (unknown) disorder parameters for each sample by comparing with the experimental density-dependent resistivity. Using these disorder fit parameters, we then obtain the effective critical density for the 2D MIT crossover to the strongly localized AI by using the well-known Ioffe-Regel-Mott (IRM) \cite{ioffeNoncrystallineAmorphousLiquid1960, mottCoulombGapLowtemperature1975} criterion for strong localization.  
A comparison between the experimental and theoretical critical density across all 2D materials shows excellent agreement, strongly suggesting that the observed 2D MIT is essentially an Anderson-like disorder-induced localization crossover rather than a Wigner-like interaction-induced crystallization, even for rather clean samples where the transition occurs at very low carrier densities. To compare and contrast with the WC transition, we follow the experimentalists, who typically claim a WC transition whenever the experimental 2D MIT occurs at a sufficiently low $n_c$ so that it is approximately consistent with the corresponding theoretical estimates based on microscopic Monte Carlo simulations \cite{ceperleyGroundStateFermion1978, lamLiquidsolidTransitionFractional1984, zhuWignerCrystallizationFractional1993, foulkesQuantumMonteCarlo2001}
for the 2D WC transition.  We find that for most materials, the theoretical WC Monte Carlo predictions occur at much lower densities than the experimentally observed $n_c$, and even for few samples where the experimental $n_c$ is consistent with the Monte Carlo WC predictions, the experimental $n_c$ is actually in better agreement with the IRM prediction for the AI transition.

We consider essentially the whole gamut of semiconductor systems experimentally reporting 2D MIT observations in transport measurements: n-GaAs \cite{lillyResistivityDilute2D2003a, sarmaTwoDimensionalMetalInsulatorTransition2005}, 
p-GaAs \cite{yoonWignerCrystallizationMetalInsulator1999, manfraTransportPercolationLowDensity2007},
Si 100 \cite{miMagnetotransportStudiesMobility2015, tracyObservationPercolationinducedTwodimensional2009, melnikovMetallicStateStrongly2020, melnikovQuantumPhaseTransition2019},
Si 111 \cite{melnikovMetallicStateStrongly2020}, 
p-Ge \cite{lodariLowPercolationDensity2021}, 
and ZnO \cite{falsonCompetingCorrelatedStates2022}.
We have actually studied many more experimental systems/samples encompassing essentially all of the published 2D MIT measurements in semiconductor layers, but the theoretical analyses for the 11 samples presented in the current work cover the entire physics rather compellingly, and presenting more results for additional samples would be an overkill with no new physics whatsoever.

The key physics underlying our rather compelling finding of all observed 2D MIT being consistent with the AI crossover, including the ones putatively claimed to be the WC transition, is the fact that the dominant disorder in semiconductor layers arises from random quenched unintentional charged impurities in the environment which interact with the carriers through exactly the same long-range Coulomb coupling as the direct electron-electron interaction does. Thus, for a fixed charged impurity disorder in a given sample, decreasing carrier density, going toward the MIT, leads invariably to enhanced impurity scattering along with increased correlation effect associated with increasing $r_s (\sim n^{-1/2})$. For a fixed 2D charged impurity density $n_\mathrm{imp}$, a possible dimensionless measure of disorder is $n_\mathrm{imp}/n$, which also increases as $n(r_s)$ decreases (increases).  In fact, for low enough $n$, the dimensionless disorder ($\sim 1/n$) must always dominate over the dimensionless interaction strength ($\sim 1/n^{-1/2}$).  Therefore, the ultimate asymptotic fate of the low density system is always an effective AI, and a WC phase can only at best be a fragile and unstable phase at some intermediate density with the high- and low-density phases always being a Fermi liquid metal and a disorder-dominated AI.  We do not think that it is possible to identify such an unstable fragile intermediate WC phase based just on dc transport measurements, since transport by itself cannot distinguish between AI and WC, both phases are localized (or pinned) insulators showing activated low-temperature transport. Thus, the claims of the observation of 2D WC based only on very low density dc transport measurements are at best speculative since the AI is always the more likely insulating phase at low carrier densities.

"Just so the readers are not misled, we emphasize that this work is not about the properties of the 2D low-density effective metallic phase ($n>n_c$), which have been extensively studied in the literature because of the interesting temperature and magnetic field dependence of the low-temperature metallic resistivity \cite{kravchenkoPossibleMetalinsulatorTransition1994a, dassarmaScreeningTransport2D2015, abrahamsMetallicBehaviorRelated2001,dassarmaSocalledTwoDimensional2005a, svkravchenkoMetalInsulatorTransition2003, spivakColloquiumTransportStrongly2010a, dassarmaElectronicTransportTwodimensional2011a}. 
There are numerous theoretical works over the last 25 years providing possible explanations for the properties of the 2D effective metallic phase, including studies emphasizing the similarity between the temperature and the magnetic field dependence of the resistivity in the metallic phase (i.e., $n>n_c$) \cite{dassarmaSimilaritiesDifferencesTwodimensional2005, fleuryEnergyScaleMetallic2010}. 
Our work is focused entirely on discerning the nature of the density-tuned 2D MIT crossover from an effectively metallic phase at higher density ($n>n_c$), where the resistivity increases with increasing temperature, to an effectively insulating phase at lower density ($n<n_c$), where the resistivity deceases with increasing temperature, and not on the detailed properties of the effective metal and the effective insulator.  In particular, we want to be able to approximately predict the values of $n_c$ for a large number of 2D systems using screened Coulomb disorder as the underlying mechanism for this metal to insulator crossover.  While much work has focused on the properties of the 2D effective metal, very little work has gone into understanding what controls the value of $n_c$, which is what we study in the current work, establishing that $n_c$ for 2D MIT is quantitatively consistent with the Anderson localization crossover in Coulomb disorder landscape."

The rest of this article is organized as follows. In Sec.~\ref{sec:2}, we provide and discuss, without any technical details, a summary of our primary findings and conclusions in order to guide the reader and as a theme for the rest of the article.  In Sec.~\ref{sec:3} we provide the basic theory. In Sec.~\ref{sec:4}, we present our results along with discussions. In Sec.~\ref{sec:5}, we conclude emphasizing our findings and suggesting future directions.

\section{Summary of the key results} \label{sec:2}
\begin{figure*}[!htb]
    \centering
    \includegraphics[width=\linewidth]{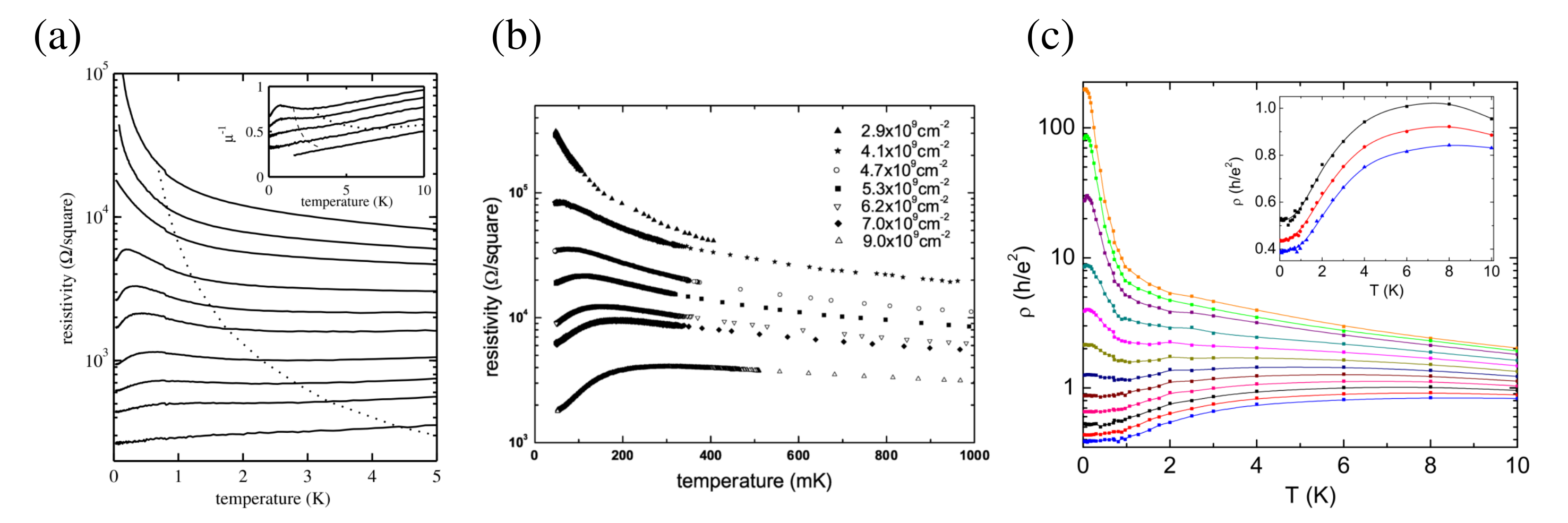}
    \caption{Measured resistivity as a function of temperature at different carrier densities for (a) nGaAs \cite{lillyResistivityDilute2D2003a}; (b) pGaAs \cite{manfraTransportPercolationLowDensity2007}; (c) nSi100\cite{tracyObservationPercolationinducedTwodimensional2009}. Each figure shows a typical MIT behavior of the temperature dependence of resistivity near the critical carrier density.  See the experimental references for the details.}  
    \label{fig:1}
\end{figure*}
In Fig.~\ref{fig:1} we show three examples \cite{lillyResistivityDilute2D2003a, manfraTransportPercolationLowDensity2007, tracyObservationPercolationinducedTwodimensional2009} 
of the typical 2D MIT data we theoretically analyze in the current work.  Figure 1 (a), (b), (c), taken from the published experimental literature, depict respectively the 2D MIT in nGaAs, pGaAs, and nSi100 systems.
   

What is plotted in all three figures is the 2D resistivity as a function of $T$ for different values of the 2D carrier density $n$. It is the low-$T$ behavior which is of interest to us for understanding MIT, and it is clear that all three figures show a transition at low $T$ from a metal (high $n$) to an insulator (low $n$), with the low-$T$ resistivity changing its behavior qualitatively. The experimental critical density, $n_\mathrm{ex}$, the key parameter of interest to us, for the MIT is the value of $n$ separating the two behaviors at $T=0$.  Obviously, $n_\mathrm{ex}$ is not known precisely, but this matters little for our analysis, and we accept the best value of $n_\mathrm{ex}$ quoted in the experiment.  The other experimental quantity of interest is the resistivity $\rho_\mathrm{ex}$ for $n=n_\mathrm{ex}$ separating the metal from the insulator: the system is a metal (insulator) for $n> (<)n_\mathrm{ex}$ and $\rho< (>)\rho_\mathrm{ex}$.  The distinction is not sharp close to $n\sim n_\mathrm{ex}$, and therefore the precise values of both $n_\mathrm{ex}$ and $\rho_\mathrm{ex}$ are approximate—in particular, $\rho_\mathrm{ex}$ could be uncertain by factors of 2-3 easily (because the resistivity varies exponentially with $T$ on the insulating side), but not by orders of magnitude.  Our theoretical task is to obtain $n_\mathrm{th}$ and $\rho_\mathrm{th}$, which are the theoretical crossover density and resistivity for the MIT, and compare them (particular the critical density) with the experimental data across all the samples involving different materials where $n_\mathrm{ex}$ varies by orders of magnitude.
Arguments claiming WC transitions manifesting as 2D MIT are based almost entirely on two premises: (1) the sample is very clean, as reflected in very high mobility (i.e. very long scattering time) deep in the metallic phase ($n>n_\mathrm{ex}$) so the correlation effects are presumably much stronger than disorder effects; and more importantly, (2) MIT happens at a very low $n_\mathrm{ex}$, where the corresponding $r_s$ value, $r_\mathrm{ex} = (n_\mathrm{ex}\pi a_B^2)^{-1/2}$, with $a_B$ being the effective Bohr radius for the appropriate material, is sufficiently large to be consistent with the estimated theoretical critical value $r_c \sim 35$ for the WC transition as deduced by the QMC numerical simulations. So far, every experiment, with the exception of one (shown in Fig.~\ref{fig:1}(b)) \cite{manfraTransportPercolationLowDensity2007}, 
where 2D MIT happened for $n_\mathrm{ex}$ low enough for the effective $r_c \sim 30$ or above, has been claimed to be an FL-WC phase transition driven by interaction effects overcoming the zero point kinetic energy leading to crystallization. The three such claims for the WC observation based on the 2D MIT critical density being low enough so that $r_\mathrm{ex} \sim r_c$ are the pGaAs system \cite{yoonWignerCrystallizationMetalInsulator1999} 
of Yoon et al., 2D ZnO system \cite{falsonCompetingCorrelatedStates2022} 
of Falson et al., and very recently, the 2D nAlAs \cite{hossainAnisotropicTwodimensionalDisordered2022}  
system, 
with $r_\mathrm{ex} \sim34$, 30, 38 respectively in these three systems
(We mention that Ref.~\cite{hossainAnisotropicTwodimensionalDisordered2022} is too recent to be included in our analysis and is cited here only for the sake of completeness although we see no reason why our general conclusion would not apply to Ref.~\cite{hossainAnisotropicTwodimensionalDisordered2022} also).
Ironically, Manfra et al.\cite{manfraTransportPercolationLowDensity2007},
with its 2D MIT behavior shown in Fig.~\ref{fig:1}(b), who reported the highest observed $r_\mathrm{ex} \sim50 (\gg r_c \sim 30)$ for any 2D MIT, did not claim a WC transition, instead attributing the metal-to-insulator crossover in their experiment in terms of a low-density disorder-driven Anderson localization transition.  This is in fact our conclusion also in the current work, where we posit the possibility that all density-tuned 2D MIT observed in transport measurements of 2D semiconductor layers (of the type presented in Fig.~\ref{fig:1}), even the ones with $r_\mathrm{ex} > r_c$, are actually disorder-driven AI transition (rather than correlation-driven WC transition) cannot be ruled out.

To buttress this claim (i.e. 2D MIT is fundamentally a disorder-driven Fermi liquid-metal to a localized-insulator transition), we provide in Table 1 the meta-summary of our theoretical analysis (presented in the next sections of this paper), which leads to our conclusion that 2D MIT manifesting in transport is dominated by disorder.

\begin{table*}[t]
	\caption{Shows a meta-summary of our theoretical analysis of 2D MIT experimental data in 11 different systems as shown in the first column. Anderson localization (Wigner crystallization) is implied by $r_\mathrm{WC}$ ($r_\mathrm{AI}$) in column 2(3) being generically closer to (or smaller than) O(1). Additional support for AI (WC) comes from $r_i$  ($r_c$) in column 4 (5) being generically of O(1) (being $\sim 36$ or larger).  The last column shows the sample quality as measured by the maximum 2D mobility (for each sample) deep in the metallic phase $n\gg n_\mathrm{ex}$. We note that the mobility is measured at different temperatures and densities for different samples. See text for details on this meta-summary. }
	\label{table:1}
	\includegraphics[width=0.8\linewidth]{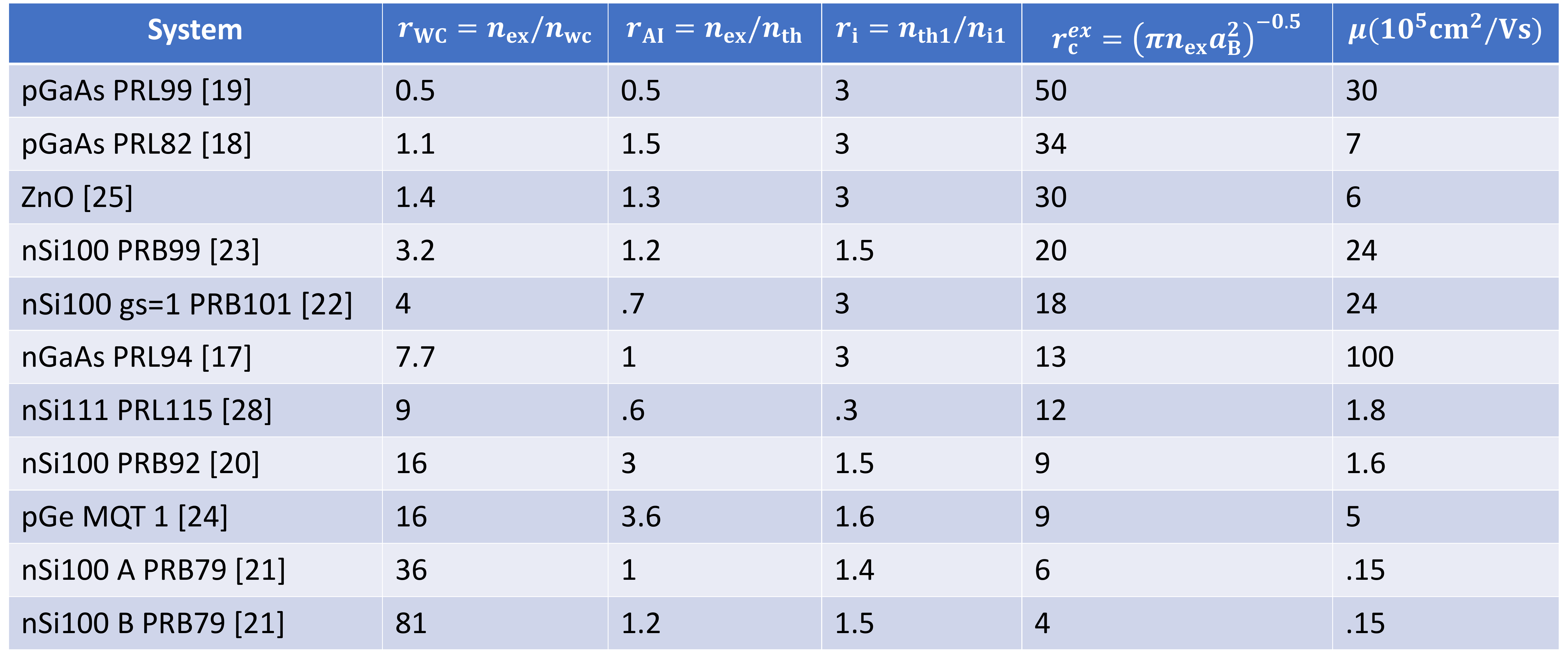}
\end{table*}


Each experimental system we analyze (three examples are in Fig.~\ref{fig:1}, and there are 8 more systems as shown in the table with appropriate references) manifests 2D MIT at a critical density $n_\mathrm{ex}$, whereas $n_\mathrm{th}$ is our theoretical prediction (as explained in the next sections of the paper) for the critical density assuming the transition to be the disorder-driven Anderson localization, with $n_\mathrm{WC}$ being the QMC theoretical prediction (assuming $r_c=36$) for the critical density to the correlation-driven WC phase. The next to the last column, $r_c^\mathrm{ex}$, shows the effective $r_s$ value for $n_\mathrm{ex}$ itself to be compared with $r_c=36$ as predicted by QMC calculations to check whether the experimental transition is occurring at or away from the predicted WC transition—closer $r_c^\mathrm{ex}$ is to 36 (particularly if it exceeds 36) the more likely it is that the transition is to a WC.  Columns 2 and 3 provide the dimensionless numbers for $r_\mathrm{WC}=n_\mathrm{ex}/n_\mathrm{WC}$ and $r_\mathrm{AI}=n_\mathrm{ex}/n_\mathrm{th}$, which respectively indicate whether the transition should be construed as a WC (if $r_\mathrm{WC}\sim O(1)$)or as an AI (if $r_\mathrm{AI} \sim O(1)$).  Finally, the column 4 is a dimensionless ratio between two theoretical densities (both explained and calculated in the next sections of the paper), where $n_\mathrm{th1}$ is the calculated theoretical value for the critical density for Anderson localization based on a minimal single-parameter disorder model, where the disorder is entirely characterized by a single random 2D charged impurity density $n_\mathrm{i1}$ in the 2D layer.  If this ratio $r_i=n_\mathrm{th1}/n_\mathrm{i1}$ is of O(1), the theoretical framework is consistent in the sense that the experimental 2D MIT is driven mainly by increasing effective disorder with decreasing carrier density rather than by increasing effective interaction with decreasing carrier density.

We emphasize that low-$T$ dc transport measurements can only distinguish between a metal and an insulator easily, efficiently, and accurately, but cannot by itself discern a WC from an AI since both are expected to manifest strongly insulating temperature dependence.  In addition, a pure WC in the presence of no disorder is in fact a super-metal as it would collectively slide in an infinitesimal applied electric field.  The insulating behavior of a WC necessarily requires the presence of disorder to pin the solid.  Thus, disorder is the essential key to the insulating behavior at low carrier density, either by causing direct Anderson localization or by indirectly pinning the WC.

The simplest theoretical idea, which is popular among the experimentalists, is to compare the experimental MIT critical density, $n_\mathrm{ex}$, to the putative QMC WC critical density $n_\mathrm{WC}$, and when the ratio $r_\mathrm{WC}=n_\mathrm{ex}/n_\mathrm{WC} <1$, the WC transition is asserted.  The column 2 in Table 1 provides $r_\mathrm{WC}$ values for the 11 systems—we emphasize that this column does not involve any theoretical work on our part.  We simply take $n_\mathrm{ex}$ from the respective experiments and choose $r_c =36$ (from QMC theories in the literature) to ascertain $n_\mathrm{WC}$. We note that column 2 indicates what we already know: Only 3 samples, two GaAs holes ($r_\mathrm{WC} =0.5$ and 1.1) and one ZnO electrons ($r_\mathrm{WC}=1.4$), have $r_\mathrm{WC} \sim 1$, and the sample with the smallest value of $r_\mathrm{WC}$ (=0.5) does not claim the observation of WC, instead analyzes its results in terms of disorder-driven strong AI localization.  All other 8 samples have $r_\mathrm{WC}\gg 1$, and cannot therefore have anything to do with a transition to the WC at the MIT, and for the three samples with $r_\mathrm{WC} \sim 1$ (or 0.5 for \cite{manfraTransportPercolationLowDensity2007}), 
the WC scenario cannot necessarily be ruled out.

Columns 3 and 4, both based on our theoretical results obtained in the current work, show that the 2D MIT in all 11 samples is consistent with Anderson localization. (This is true, independent of the specific value of $r_\mathrm{WC}$ given in column 2 whether it is of O(1) or O($\gg $1).)  Column 2, a key finding of our current detailed meta-analysis of all existing 2D MIT experiments, shows that the experimentally extracted effective crossover density $n_\mathrm{ex}$ for all samples is always consistent with the 2D MIT being a crossover from an effective metal (for $n>n_\mathrm{ex}$) to an effective strongly localized AI (for $n<n_\mathrm{ex}$).  This is clearly seen in the dimensionless ratio $r_\mathrm{AI} =n_\mathrm{ex}/n_\mathrm{th}$ being of O(1) for all samples independent of whether $n_\mathrm{ex}$ is much higher than or even lower than $n_\mathrm{WC}$, the QM-predicted WC transition density.  Here $n_\mathrm{th}$ is our calculated sample-dependent critical density for the AI transition for each sample, which strongly depends on the details of each sample, 
varying from $n_\mathrm{th}\sim 10^{11} \mathrm{cm}^{-2}$ for dirty Si samples to $n_\mathrm{th}\sim 10^{10} \mathrm{cm}^{-2}$ for Ge, all the way to $n_\mathrm{th}\sim 10^9 \mathrm{cm}^-2$ for the ultraclean n-GaAs sample.  
In addition, column 4 shows that a single-disorder parameter approximate theory always gives the dimensionless disorder parameter $r_i = n_\mathrm{th1}/n_\mathrm{i1}$ to be of O(1) for all samples, where $n_\mathrm{th1}$ is the theoretically calculated AI crossover density and $n_\mathrm{i1}$ is the necessary best fit value of the disorder parameter $n_\mathrm{i1}$ needed to quantitatively describe the metallic transport data ($n>n_\mathrm{ex}$).  Essentially, columns 2-4 together make a compelling case for all observed 2D MIT being disorder-dominated Anderson localization crossover with the observed critical MIT density $n_\mathrm{ex}$ being very low in the cleanest samples simply by virtue of the relevant impurity density being very low.  This very low $n_\mathrm{ex}$ in clean samples may accidentally be comparable to the corresponding value of the QMC-predicted WC transition density (as it is for 3 samples in Table \ref{table:1}), but no particular significance can be attached to this coincidence since columns 3 and 4 strongly suggest that the MIT, even in these clean samples, is an AI transition, as was already concluded in Ref. \cite{manfraTransportPercolationLowDensity2007}, 
which manifests by far the largest effective value of critical $r_s (>36)$ for the 2D MIT.

The column 5 in Table \ref{table:1} provides the values of $n_\mathrm{ex}$ in units of effective Bohr radii in each sample so that one can see that the effective critical sample dependent $r_s$ value coincidentally lies close to the QMC WC value 36 for 3 samples (2 pGaAs and 1 nZnO).  The last column shows the measured high-density mobility deep in the metallic phase ($n\gg n_\mathrm{ex}$), providing a rough measure of the sample quality or cleanliness. One interesting fact reflected in this column 6 is that the mobility in the three samples where $r_c$ (column 2) $\sim$ O(1) is by no means spectacularly high—in fact the highest mobility sample by far \cite{lillyResistivityDilute2D2003a} 
$\sim 10^7 \mathrm{cm}^2/Vs$ has a modest $r_\mathrm{WC} \sim 7.7$ whereas the nZnO sample, which has a modest mobility $\sim 6\times 10^5 \mathrm{cm}^2/Vs$, has $r_\mathrm{WC} = 1.4$.  

In the next two sections, we describe and discuss our theory and results for 2D MIT underlying the meta-summary presented in Table \ref{table:1}.

\section{Theory} \label{sec:3}
The basic theory has two parts.  In the first part, which is laborious, we develop a theory for impurity scattering low-T transport in 2D semiconductors, and vary the impurity disorder parameter to obtain the best possible fits to the density-dependent resistivity of each sample for the whole range of density over which the experimental results are available always staying in the metallic regime $n > n_\mathrm{ex}$. In order to avoid any unnecessary complications arising from the temperature dependence of the resistivity, we consider only the lowest-$T$ experimental data at the base temperature and compare it to our calculated $T=0$ density-dependent resistivity. The temperature-dependent 2D resistivity itself is a subject of considerable interest and importance \cite{dassarmaScreeningTransport2D2015}, but it is beyond the scope of the current work where we focus on the MIT itself which is an effective T=0 crossover, so our interest is on how the resistivity changes as a function of carrier density at $T=0$ in going from the metal ($n>n_\mathrm{ex}$) to the insulator ($n<n_\mathrm{ex}$).
We calculate the density-dependent metallic resistivity, $\rho(n)$, at $T=0$ as a function of disorder parameters in the 2D sample by calculating the scattering time or the relaxation time $\tau (k, n)$, given by 
\begin{align} \label{eq:1}
    \frac{1}{\tau(k, n)}&=\frac{2\pi}{\hbar} \int dz  
    N_i(z) 
    \sum_{\bm k'}
    \left|u_{\bm k - \bm k'}(z)\right|^2 \\ \nonumber
    &\times
    (1-\cos{ \theta_{\bm k, \bm k'}})\delta(\epsilon_{\bm k}-\epsilon_{\bm k'}),
\end{align}
where $\epsilon_{\bm k}=\hbar^2k^2/2m$ is the usual parabolic energy dispersion with $k=|\bm k|$, $\theta_{\bm k- \bm k'}$ is the angle between the wavevector of the incoming ($\bm k$) and outgoing ($\bm k'$) states, and $N_i(z)$ denotes the disorder defined by a random 3D charged impurity density along the $z$-direction which is normal to the 2D layer along the $x$-$y$ plane located at $z=0$. The scattering time has no directional dependence due to the rotational symmetry of the energy dispersion, and thus can be written as $\tau(k, n)$ instead of $\tau(\bm k, n)$.  In Eq.~(\ref{eq:1}), $u_{\bm k - \bm k'}$ is the screened disorder potential arising from the Coulomb interaction between the 2D carriers and the random charged impurities in the environment, which is described in more detail in Eq.~(\ref{eq:2}).  Eq.~(\ref{eq:1}) defines the well-established and extremely successful time-tested minimal metallic transport model for 2D \cite{andoElectronicPropertiesTwodimensional1982, dassarmaScreeningTransport2D2015}.
Note that we ignore all complications of weak localization in Eq.~(\ref{eq:1}) which is semi-classical in nature-- this is simply because of the fact that at high density ($n>n_c$),  Eq.~(\ref{eq:1}) is quantitatively valid since weak localization corrections are negligibly small showing up as very weak logarithmic effects at very low temperatures \cite{dassarmaSignaturesLocalizationEffective2014}. 

At $T=0$, the only relevant scattering is at the Fermi surface, $q=k_\mathrm{F}$ and we define $\tau(q=k_\mathrm{F})=\tau$.  Note that Eq.~(\ref{eq:1}) simplifies for $T=0$ since the delta function associated with the energy conservation guarantees that the momentum transfer insider the integral cannot exceed 2$k_\mathrm{F}$.
The screened disorder potential $u_{\bm q}(z)$ is given by:
\begin{equation} \label{eq:2}
    u_{\bm q}(z) = \frac{v_{\bm q}}{ \varepsilon(q)}e^{-qz}= \frac{2\pi e^2}{\varepsilon(q) \kappa q } e^{-qz}.
\end{equation}
Here $v_{\bm q}$ is the 2D Coulomb coupling in momentum space, $v_q =2\pi e^2/\kappa q$, with $\kappa$ as the background lattice dielectric constant (typically, $\kappa \sim 10$, but it varies with the semiconductor).  The exponential factor in Eq.~(\ref{eq:2}) arises from the charged impurities being quenched in the 3D environment around the 2D semiconductor layer where the electrons are confined in the $z=0$ plane.
In Eq.~(\ref{eq:2}), the dielectric (screening) function introduces strong carrier density dependence in the effective (screened) disorder since the 2D polarizability function $\Pi(\bm q)$ depends on the 2D carriers themselves, given by:
\begin{align}     \label{eq:3}
    \Pi(\bm q)=-\frac{m}{\pi \hbar^2}
		\left[1 - \Theta(q-2k_\mathrm{F})\frac{\sqrt{q^2- 4k^2_\mathrm{F} }}{q} \right].
\end{align}
Here, $m$ is the system-dependent carrier effective mass, and $k_\mathrm{F}$ is the density-dependent 2D Fermi momentum defined by:
\begin{equation} \label{eq:4}
    k_\mathrm{F} = \sqrt{\frac{2\pi n}{g_v}}
\end{equation}
A spin degeneracy of 2 is assumed in Eq.~(\ref{eq:4}) and $g_v$ is the system dependent valley degeneracy with $g_v= 1$ (p-Ge, p- and n-GaAs, and ZnO), 2 (n-Si 100), 6 (n-Si 111) for the systems under consideration.  If the spin degeneracy is lifted (as in one sample we consider), the factor of 2 in Eq.~(\ref{eq:4}) is replaced by 4.
Inserting the 2D polarizability function in the dielectric screening function, we find that the screened disorder potential in Eq.~(\ref{eq:1}) is given by the following simple formula:
\begin{equation} \label{eq:5}
    u_{\bm q}(z) =  \frac{2\pi e^2}{\kappa (q + q_\mathrm{TF}) } e^{-qz}.
\end{equation}
Here the Thomas-Fermi screening wave-vector $q_\mathrm{TF}$ suppressing the long wavelength $q=0$ divergence of the Coulomb coupling is defined as
\begin{equation} \label{eq:6}
    q_\mathrm{TF}=\frac{2me^2g_v}{\kappa \hbar^2}
\end{equation}

Again, a spin degeneracy $g_s=2$ is assumed in Eq.~(\ref{eq:6})—for $g_s=1$ the factor of 2 becomes just unity in Eq.~(\ref{eq:6}).
We note that the $T=0$ transport theory simplifies with the screened disorder $u (q)$ given simply by Eq.~(\ref{eq:5}) because the energy conservation associated with the elastic impurity scattering assures that all scattering wave-vector $q<2k_\mathrm{F}$.  2D screening is momentum-independent for $q=0$ to $2k_\mathrm{F}$ by virtue of the 2D density of states being a constant.  All we need is $\tau = \tau (k_\mathrm{F})$
We see, from the theory described above, that the system and sample details enter transport through 4 parameters: $g_s$, $g_v$, $\kappa$, $m$. Carrier density enters through $k_\mathrm{F}$.
All one needs now is the impurity distribution function $N_i(z)$ and then Eq.~(\ref{eq:1}) directly gives the scattering time for the resistivity. Unfortunately, we face a serious problem at this point because, by definition, the details of charged impurity disorder are unknown since the charged impurities are unintentional, and little independent information is available about them. In fact, transport data are invariably the best source of information for the impurity disorder since the experimental density-dependent resistivity can be compared with the transport theory results to derive the impurity distribution. This is precisely what we do for each sample.

Since the 3D disorder distribution $N_i(z)$ is, in principle, defined by an infinite number of parameters, an extensively used convenient model is a simple 2-parameter ($n_i$ and $d$ in Eq. 7 below) effective disorder model, where the full impurity distribution is replaced by:
\begin{equation} \label{eq:7}
    N_i(z)=n_i\delta(z-d),
\end{equation}
where a 2D charged impurity layer with impurity density $n_i$ is placed at a distance $d$ from the 2D layer.  The model simplifies even further into a 1-parameter model if $d=0$ is used with the charged impurities of 2D density $n_\mathrm{i1}$ placed in the same plane ($z=0$) as the 2D layer itself.
\begin{equation} \label{eq:8}
N_i(z)=n_\mathrm{i1} \delta(z)
\end{equation}

We utilize both of these 2- and 1-parameter models in our theory, tuning the disorder parameters to produce the best overall agreement between the theory and the experimental transport data to extract $n_i$ and $d$.  We emphasize that we do not allow $n_i$ and $d$ to be carrier density dependent as one may typically do in device simulations.  Our extracted disorder parameters vary from sample to sample (and cleaner samples with higher mobility typically have smaller/larger $n_i/d$), but for a given sample, the disorder parameters are fixed and do not vary, so that all the variations in the resistivity as a function of carrier density in a given sample arise only from the intrinsic effects of the carrier density. The single-parameter disorder model obviously provides poor fits to the data because of its physical inapplicability to 2D semiconductors where the dominant disorder is never in the 2D layer itself.
Once the scattering time is calculated, the resistivity is given by the standard T=0 Drude formula:
\begin{equation} \label{eq:9}
    \rho = \frac{m}{ne\tau^2}.
\end{equation}
We note that $n$ and $m$ enter again in the calculation of the resistivity (in addition to the implicit dependence of $\tau$ itself on $n$ and $m$ through Eqs.~(\ref{eq:1})-(\ref{eq:7})). Thus, even within our minimal transport model, the resistivity depends on 7 independent parameters $n$, $m$, $g_s$, $g_v$, $\kappa$, $n_i$, $d$ in a complex manner, and such a high-dimensional parameter dependence cannot be described by any simplistic considerations.

The first part of our theory described above involves calculating $\rho(n)$ for each sample as a function of the disorder parameters $n_i$ and $d$, and then obtaining the recursive best fit of the experimental resistivity at the lowest $T$ to the theoretical results over the whole metallic density regime ($n>n_\mathrm{ex}$).  These recursive fits provide two disorder parameters $n_i$ and $d$ (or just one parameter $n_\mathrm{i1}$) as described in Eq.~\ref{eq:7} (\ref{eq:8}) for 2-parameter (1-parameter) disorder models. Although the 1-parameter fit is by definition the minimal mathematical model for disorder, the 2-parameter fit is the minimal physical model describing the experimental data much better.  We do, however, carry out extensive fitting using the 1-parameter model also because the 1-paramater model enables a direct consistent microscopic distinction between the strong- or weak- disorder regime based on whether the extracted impurity density parameter $n_\mathrm{i1}$ is larger or smaller than the relevant carrier density $n$ itself.

Once the disorder parameters are extracted by our recursive fits of the experimental data, we calculate the theoretical critical density $n_\mathrm{IRM}$ (which we refer to as $n_\mathrm{th}$, implying that it is our best theoretical estimate for the critical density). This of course explicitly assumes the MIT to be an Anderson localization crossover, with the system being a metal (Anderson insulator) for $n > (<) n_\mathrm{IRM}$, by using the Ioffe-Regel-Mott criterion:
\begin{equation} \label{eq:10}
    E_\mathrm{F}\tau = \hbar.
\end{equation}
Here $\tau$ is the scattering time of Eq.~(\ref{eq:1}), which depends on the system parameters (e.g., $m$, $g_s$, $g_v$, $\kappa$), the carrier density $n$, and the disorder parameters (e.g. $n_i/d$ or $n_\mathrm{i1}$).  The Fermi energy $E_\mathrm{F}$ is given by 
\begin{align} \label{eq:11}
    E_\mathrm{F} = \frac{\hbar^2k_\mathrm{F}^2}{2m} = \frac{\hbar^2}{2m}\frac{4\pi n}{g_sg_v}
\end{align}
Writing $\Gamma =\hbar/\tau$, Eq.(10) is sometimes rewritten as
\begin{equation} \label{eq:12}
    E_\mathrm{F}=\Gamma
\end{equation}
where $\Gamma$ is the incoherent (level) broadening arising from impurity scattering, and Eq.~(\ref{eq:12}) or (\ref{eq:10}) simply implies that coherent or metallic transport carried by well-defined quasiparticles becomes impossible once the density (disorder) is low (high) enough to satisfy Eq.~(\ref{eq:10})/(\ref{eq:12}).  This condition is precisely equivalent to the better-known IRM criterion written as $k_\mathrm{F} l=1$, where $l$ is the mean free path given by $l=v_\mathrm{F} \tau$ with $v_\mathrm{F}=\hbar k_\mathrm{F}/m$, the Fermi velocity.

We note that one immediate theoretical consequence (obtained by combining Eqs.~(\ref{eq:9})-(\ref{eq:11})) of the IRM criterion is that the theoretical 2D resistivity at the MIT crossover is given by the simple formula:
\begin{equation} \label{eq:13}
    \rho_\mathrm{IRM}= \frac{2}{g_s g_v}\frac{h}{e^2}.
\end{equation}
Although one should not take this crossover resistivity as a sharply defined critical resistivity, the 2D AI transition should generically show a crossover resistivity $\sim h/e^2\sim 26$ k-ohms for $g_s=2$ and $g_v=1$ with the metallic (insulating) phase typically having resistivity below (above) 26 k-ohms provided the MIT is an AI transition.  No such special crossover resistivity value should however characterize a WC-based 2D MIT.

Using the best-fit disorder parameters obtained (in the first part of our theoretical work) for each sample by analyzing the experimental data in the metallic ($n>n_\mathrm{ex}$) regime, we then solve the integral equation defined by the IRM criterion in Eq.~(\ref{eq:10}) for the Anderson localization crossover (critical) theoretical density $n_\mathrm{IRM}$.  We emphasize that our calculated $n_\mathrm{IRM}$ provides the theoretical estimate for the 2D MIT transition $n_\mathrm{ex}$ explicitly assuming the MIT to be Anderson localization, with no consideration of Wigner localization whatsoever. The theory uses the IRM criterion for the AI crossover and the underlying physics driving the MIT is exclusively random disorder in the environment (and not electron correlation effects which are responsible for the WC transition from a FL metal to a WC insulator).
We discuss our results for all 11 samples in the next section.

\section{Results} \label{sec:4}
\begin{figure*}[!htb]
    \centering
    \includegraphics[width=\linewidth]{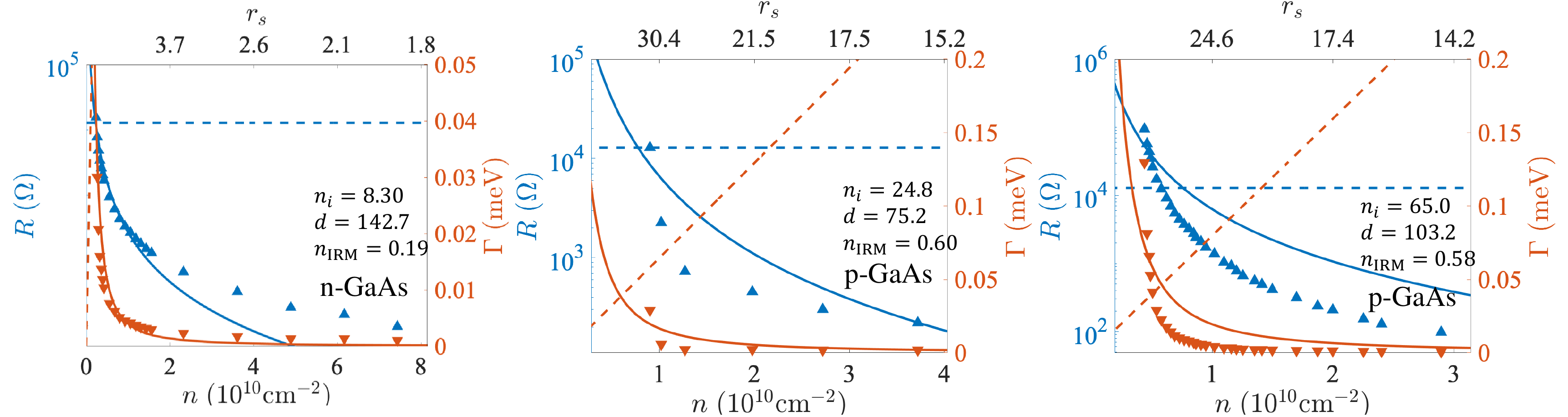}
    \caption{Shows the experimental data and our best fit theory along with the theoretically estimated disorder parameters and 2D MIT density using the IRM criterion for (a) n-GaAs \cite{sarmaTwoDimensionalMetalInsulatorTransition2005} [Phys. Rev. Lett 94, 136401]; (b) p-GaAs \cite{yoonWignerCrystallizationMetalInsulator1999} [Phys. Rev. Lett. 82, 1744]; (c) p-GaAs \cite{manfraTransportPercolationLowDensity2007} [Phys. Rev. Lett. 99, 236402]. Red (blue) up (down) triangles are the experimental data for the resistivity $\rho$ and the incoherent broadening $\Gamma$ respectively, and the solid lines are the best theoretical recursive fits in the whole metallic density regime.  The red dashed lines are the Fermi energy $E_\mathrm{F}$. The IRM condition $\Gamma = E_\mathrm{F}$ defines the 2D MIT crossover point. Both carrier density $n$ and the corresponding dimensionless interaction parameter $r_s$ are shown in each panel. Units for $n_\mathrm{i}/n_\mathrm{IRM}$ and $d$ are $10^{10} \mathrm{cm}^{-2}$ and nm respectively.}  
    \label{fig:2}
\end{figure*}

\begin{figure*}[!htb]
    \centering
    \includegraphics[width=\linewidth]{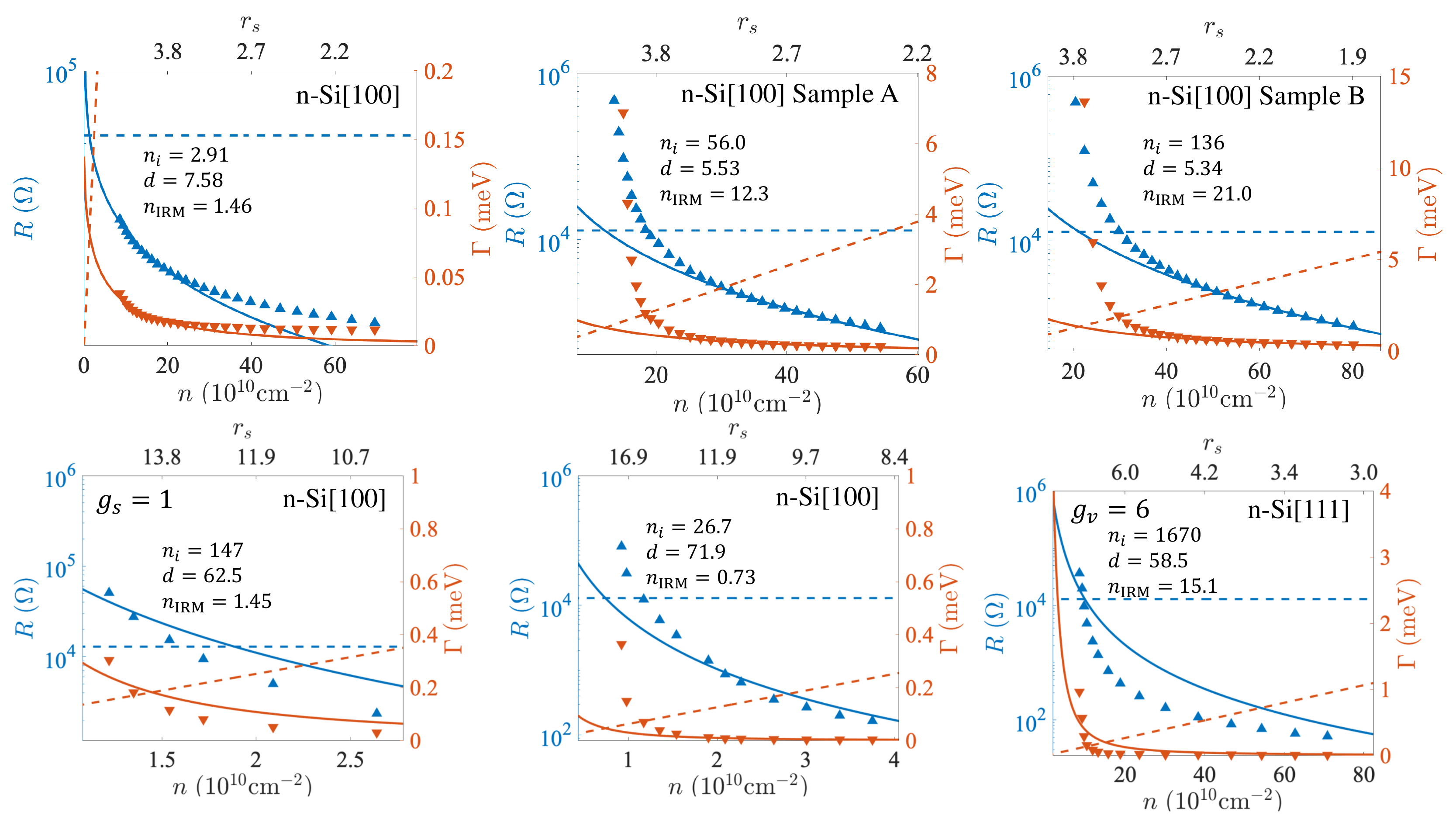}
    \caption{Shows the same as in Fig.~\ref{fig:2} for 6 different 2D Si samples: 
    (a) , 
    Phys. Rev. B 92, 035304 \cite{miMagnetotransportStudiesMobility2015}; 
    (b) Phys. Rev. B 79, 235307 \cite{tracyObservationPercolationinducedTwodimensional2009}; 
    (c) Phys. Rev. B 79, 235307 \cite{tracyObservationPercolationinducedTwodimensional2009}; 
    (d) Phys. Rev. B 101, 045302 \cite{melnikovMetallicStateStrongly2020}; 
    (e) Phys. Rev. B 99, 081106 \cite{melnikovQuantumPhaseTransition2019}; 
    and (f) PRL 115, 036801 \cite{huStronglyMetallicElectron2015}
    }.  
    \label{fig:3}
\end{figure*}

\begin{figure}[!htb]
    \centering
    \includegraphics[width=\linewidth]{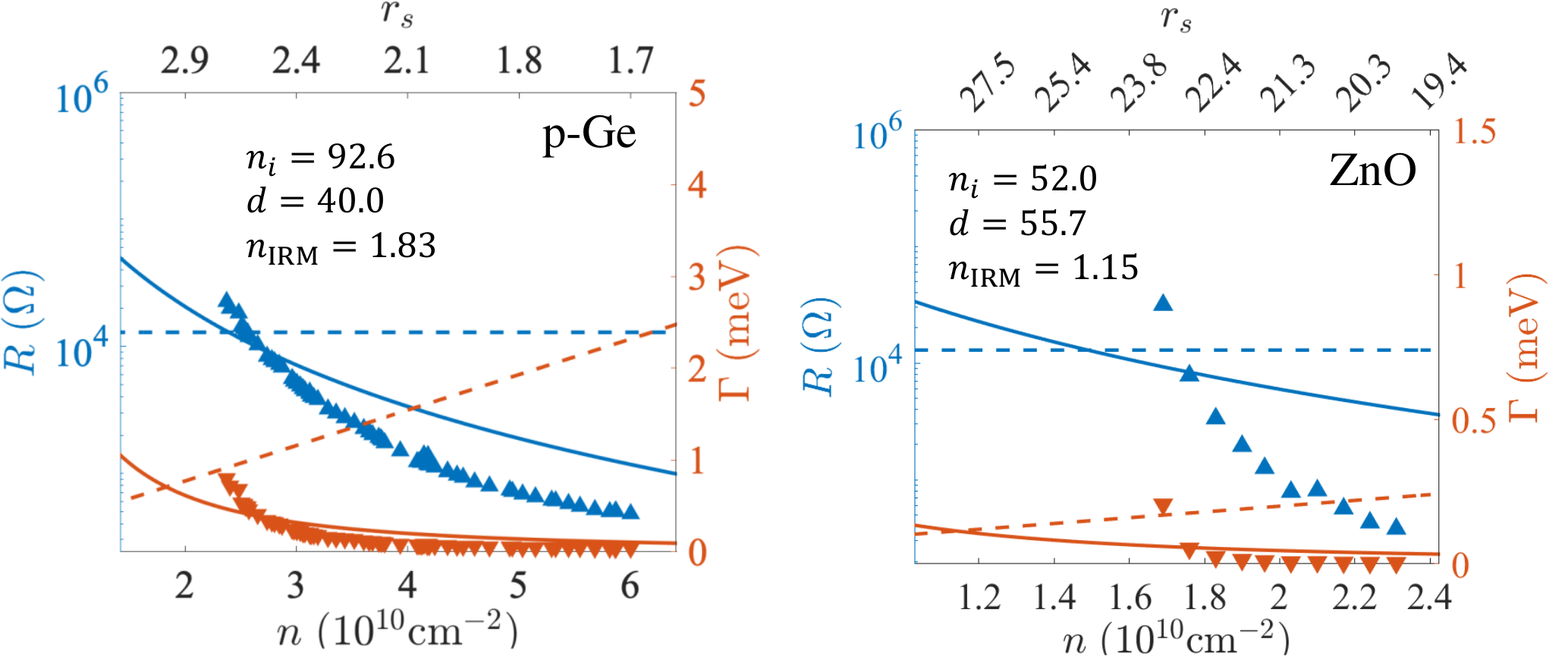}
    \caption{Fig.~\ref{fig:4}: Shows the same as in Figs. 2 and 3 for: (a) 2D p-Ge \cite{lodariLowPercolationDensity2021} [Mater. Quantum Technol. 1, 011002]; (b) n-ZnO \cite{falsonCompetingCorrelatedStates2022} [Nat Materials 21, 311].  
    }  
    \label{fig:4}
\end{figure}

We show in Figs. \ref{fig:2}-\ref{fig:4} our density-dependent resistivity results in 11 different panels (3 panels in Fig.~\ref{fig:2} for 2D GaAs, 6 panels in Fig.~\ref{fig:3} for Si samples, and 2 panels in Fig.~\ref{fig:4} for Ge and ZnO), combining the results of both parts of our theory (fitting our 2-parameter transport theory to the experimental metallic resistivity and then using the extracted disorder parameters to obtain the critical density for 2D MIT).

Before discussing our 2D MIT results, it is important to emphasize, in the context of the 11 sets of results presented in Figs. \ref{fig:2}-\ref{fig:4}, that our theory obviously does not do an equally good job in quantitatively describing all the data in all the samples—in fact, for some samples, (e.g. ZnO in Fig.~\ref{fig:4}(b)) the theory is simply poor.  This is a combined reflection of two completely different problems: (1) our two-parameter disorder model including only random charged impurity scattering could be a very poor approximation for specific samples; (2) the theory itself is approximate as it is based explicitly on the Boltzmann-RPA transport theory within the relaxation time approximation, which could break down for specific samples, particularly at low carrier densities. Unfortunately, neither problem allows for any systematic resolution, and it is essentially impossible to go beyond the approximation schemes used in the current work.  For the first problem, the background disorder details are unknown in the experimental samples, and forcing agreement with the experimental data by simply introducing a large number of disorder parameters in the theory is a meaningless empirical data-fitting procedure which would shed little light on the underlying physics.  For the second problem, we know of no systematic quantitatively predictive transport theory in strongly disordered and strongly interacting electron systems, and the theory used in the current work is essentially exact in the high-density metallic regime (except for weak localization corrections, which are not of much significance here).  Indeed, if we restrict the theory deep into the metallic phase at very high carrier densities, we can obtain essentially exact agreement between theory and experiments (except for a few samples with possible pathological disorder distributions), but such a high-density description does not extrapolate well to the low carrier densities, and hence the effective disorder parameters obtained from only a high-density fit to the data in general does not do as good a job of describing the experimental 2D MIT the whole-density-range approximate fits used in Figs. \ref{fig:2}-\ref{fig:4}. The RPA-Boltzmann transport theory is an effective mean field kinetic theory, which provides an excellent qualitative and semi-quantitative description of the metallic transport properties as long as the system can be described in terms of quasiparticles scattering off impurities, i.e., as long as it is in an effective metallic phase ($n>n_\mathrm{IRM}$).  The theory becomes progressively less accurate quantitatively as the 2D MIT is approached from the metallic side.  Obviously, the theory fails completely for $n < n_\mathrm{IRM}$, by definition, but should remain valid qualitatively all the way down to $n_\mathrm{IRM}$ where the quasiparticle kinetic description presumably no longer applies as the system has lost all coherence with no Fermi surface in the momentum space.  Until the MIT occurs, however, no part of the RPA-Boltzmann transport theory manifests any pathology except for the carrier mean free path smoothly and continuously decreasing in magnitude and passing through $1/k_\mathrm{F}$ at the crossover.  In fact, the theory itself remains perfectly mathematically consistent independent of how large (small) the broadening (mean free path) becomes except the physical basis of its foundation on the existence of well-defined quasiparticles is no longer valid for $\Gamma > E_\mathrm{F}$.
We emphasize, however, that the theory is sometimes only in qualitative (or at best semiquantitative) agreement with the experimental data (as for Fig.~\ref{fig:3} and a few other places), and this is expected of a single parameter first principles calculation.  The important point is that even for these quantitatively inaccurate situations, it seems clear that the AI description of the 2D MIT is far superior to the WC description, which is our central qualitative claim.

We emphasize that although it may not be obvious from a visual inspection of the results in Figs.~\ref{fig:2}-\ref{fig:4}, our fits of the 2-parameter theory to the experimental resistivity is indeed a recursive least squares fit over the whole available metallic density regime, leading to the estimates for the two disorder parameters $n_\mathrm{i}$ and $d$. We have also carried out a 1-parameter fit theory (taking $d=0$), which is generically far worse than the 2-parameter fit results shown in Figs. \ref{fig:2}-\ref{fig:4}. The 1-parameter fit produces a single disorder parameter $n_\mathrm{i1}$ which should be construed as the impurity density right in 2D electron layer itself, and then use of this $n_\mathrm{i1}$ enables a cruder estimate for the 2D MIT crossover density, which we call $n_\mathrm{1th}$.  We thus have two disorder models and two estimates for the 2D MIT crossover density: 2-parameter model ($n_\mathrm{i}$ and $d$) and 1-parameter model ($n_\mathrm{1i}$) with the MIT crossover density $n_\mathrm{IRM}$ and $n_\mathrm{1th}$, respectively.  These calculated theoretical crossover densities for the AI transition (as defined by the IRM criterion) are to be compared with the experimentally measured 2D MIT crossover density $\sim n_\mathrm{ex}$.
In Table \ref{table:2}, we provide a summary of our results for all 11 samples along with the relevant experimental results as well as several other important quantities for a detailed comparison between theory and experiment as well for coming to an informed conclusion about the nature of the 2D MIT phenomenon (i.e. AI or WC).

\begin{table*}[t]
	\caption{Theoretical and experimental numbers of relevance to 2D MIT in eleven 2D samples (Columns 1 and 2); column 3 shows the experimental mobility (in units of $10^5 \mathrm{cm}^2/Vs$) deep into the metallic phase indicating sample quality; column 4 gives the experimental MIT density to be compared with the theoretically derived disorder-driven critical density for the AI transition in column 5 for the 2-parameter disorder model; column 6 gives the QMC-derived theoretical correlation-driven transition density to a WC; column 7 provides the theoretical percolation density for the classical percolation MIT in the 2-parameter disorder model (for the 1-parameter model, the percolation density is the same as the impurity density itself); columns 8 and 9 provide respectively the experimental and theoretical critical resistance at the MIT; columns 10 and 11 give the best-fit disorder parameters for the 2-parameter disorder model; columns 12 and 13 give respectively the calculated MIT crossover density and the impurity density  for the 1-parameter model; column 14 gives the effective 3D impurity density, $n_3=n_i/d$ in the 2-parameter model; column 15 gives the effective interaction $r_s$ parameter value for the experimental critical density (to be compared with the corresponding QMC prediction for 2D WC transition at $r_s=36$); column 16 gives the ratio of $n_\mathrm{i1}$  (column 12) to $n_\mathrm{th1}$ (column 13) in the 1-parameter model. All densities are in units of $10^{10} \mathrm{cm}^{-2}$ and all resistance in units of $10^4$ ohms.}
	\label{table:2}
	\includegraphics[width=0.8\linewidth]{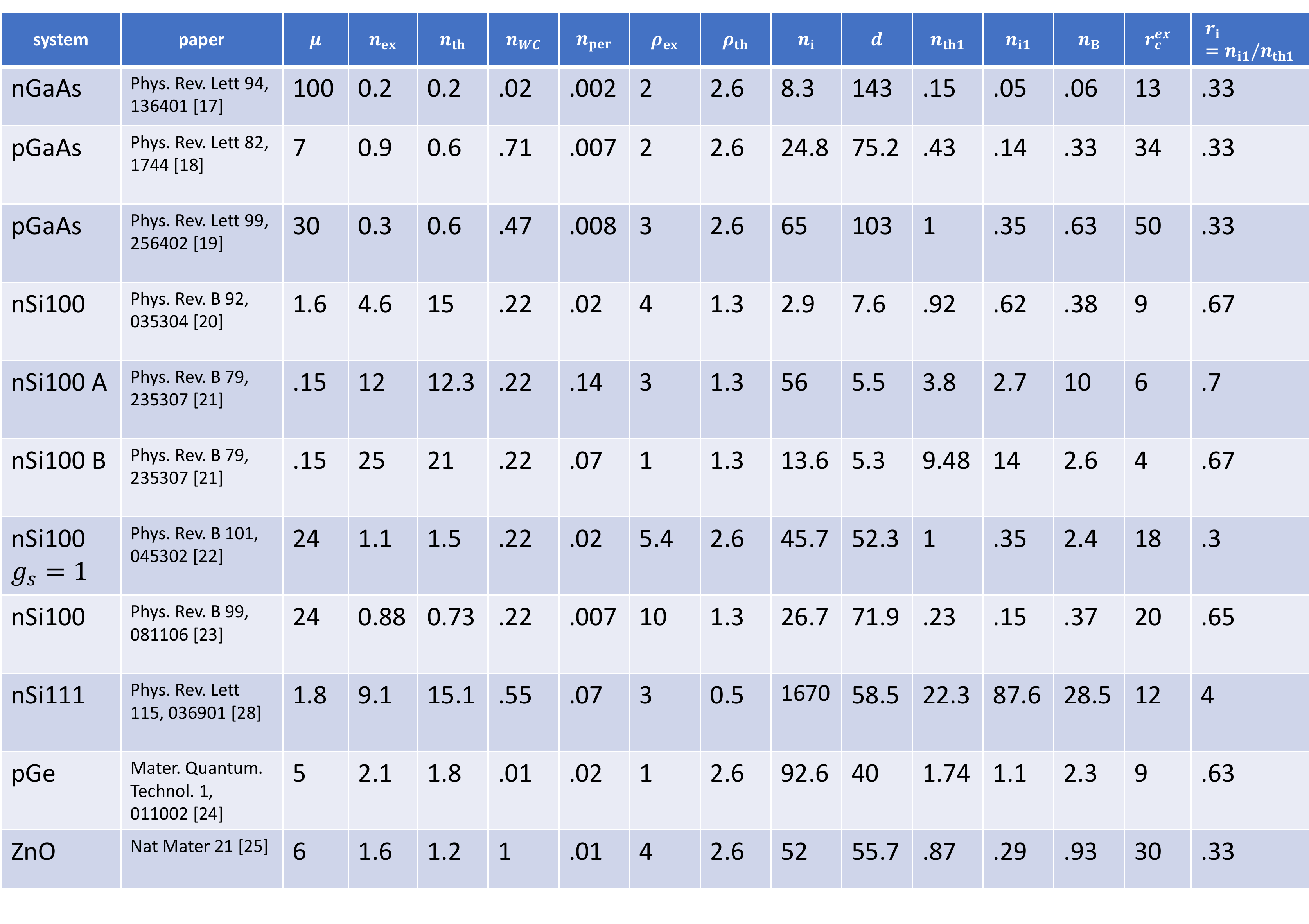}
\end{table*}

It is clear, from the extensive numbers and comparisons provided in Table \ref{table:2}, that all 2D MIT experimental observations are consistent with disorder-driven AI crossover, and most are inconsistent with correlation-driven WC crossover. Since the experimental signatures and characteristics of the 2D MIT phenomenology are essentially identical in all the 11 representative samples we study, we conclude that 2D MIT is a primarily a disorder-driven Anderson localization crossover rather than an interaction-driven Wigner crystallization, even if in a few particularly clean samples, the crossover MIT critical density just happens to be low enough coincidentally to be consistent with the QMC-predicted WC transition density.

We now elaborate on the salient features of Table \ref{table:2} (and by extension, of Figs. \ref{fig:2}-\ref{fig:4}) in the context of the underlying physics controlling 2D MIT.

(1) The most important finding in Table \ref{table:2} is that $n_\mathrm{ex}/n_\mathrm{th}$ is of O(1) essentially for all the samples with the exceptions of two samples \cite{miMagnetotransportStudiesMobility2015, lodariLowPercolationDensity2021},
where this ratio is of O(3).  But this discrepancy arises in these two samples because the experiments were done here at the relatively high $T\sim 1$K rather than in a dilution fridge with $T <100$ mK as in all other samples.  The experimental MIT critical density $n_\mathrm{ex}$ decreases with decreasing T because of the T-dependence of the metallic resistivity at lower densities (this is obvious in Fig.~\ref{fig:1}), and we believe that the true $T=0$ $n_\mathrm{ex}$ values in these two samples are much smaller than the ones reported experimentally, which would imply that $n_\mathrm{ex}/n_\mathrm{th}$ is actually of O(1) in these two cases too.  The consistent agreement between our theoretically calculated AI transition density and the experimentally observed MIT critical density strongly argues for the scenario that the 2D MIT is a generic disorder-induced AI transition, independent of how low $n_\mathrm{ex}$ might be.

(2) In looking for consistency with the WC scenario for 2D MIT, we follow the main argument of the experimentalists where any consistency between $n_\mathrm{ex}$ and $n_\mathrm{wc}$ (the putative QMC predicted critical density in the relevant sample for the Fermi liquid to the WC transition, assuming it happens at $r_s \sim 36$) is taken to be evidence supporting the WC transition. In Table \ref{table:2}, $n_\mathrm{ex} \gg  n_\mathrm{wc}$ for all, except for 3 samples: $n_\mathrm{ex}/n_\mathrm{wc} \sim 0.64$ (p-GaAs 99 \cite{manfraTransportPercolationLowDensity2007}); 1.3 (p-GaAs 82 \cite{miMagnetotransportStudiesMobility2015}); 1.6 (ZnO \cite{falsonCompetingCorrelatedStates2022}).  For the other 8 systems, $n_\mathrm{ex}/n_\mathrm{wc} \gg 1$, and therefore the WC scenario simply is not relevant at all. Of the three samples where $n_\mathrm{ex}/n_\mathrm{wc} \sim$ O(1), only one, pGaAs 99, has $n_\mathrm{ex}<n_\mathrm{wc}$, and this particular work concluded that the 2D MIT observed therein is a strong localization transition, not Wigner crystallization. One can therefore question the claims in Refs.~\cite{yoonWignerCrystallizationMetalInsulator1999, falsonCompetingCorrelatedStates2022} 
that the 2D MIT reported in these works are decisively the observation of Wigner crystallization.  The current work indicates that the experiments are more consistently interpreted as low-density crossovers to the AI phase because of low disorder in these clean systems. The fact that the critical MIT density is consistent with the Wigner crystallization theoretical prediction may simply  be a coincidence  in these particularly clean samples.

(3) The 1-parameter disorder model, while being quantitatively not very accurate, provides important insight into the physics of 2D MIT. First, we note that for most samples the 1-parameter model gives an $n_\mathrm{IRM} =n_\mathrm{th1}$ which is comparable to the result $n_\mathrm{th}$ for $n_\mathrm{IRM}$ obtained in the 2-parameter disorder model, i.e., $n_\mathrm{th1}/n_\mathrm{th} \sim O(1)$.  But the significant point about the 1-parameter disorder model is that the extracted impurity parameter $n_\mathrm{i1}$ is comparable to the calculated MIT critical density nth1 for essentially all the samples, $n_\mathrm{th1}/n_\mathrm{i1}\sim$ O(1-3).  This means that the 2D MIT happens when the sample carrier density roughly equals the putative sample impurity density, and not when the carrier density roughly equals the putative QCM WC theoretical critical density (except for some sheer coincidences).

(4) We note as a minor point of interest that the approximate 2D percolation density $n_\mathrm{per} \sim 0.1 \sqrt{n_i/d}$ \cite{efrosMetalnonmetalTransitionHeterostructures1989, nixonPotentialFluctuationsHeterostructure1990, efrosDensityStatesTwodimensional1993, dassarmaTwodimensionalMetalinsulatorTransition2013}
in Table \ref{table:2} obtained on the basis of our 2-parameter extracted impurity parameters is always smaller than the experimental critical density $n_\mathrm{ex}$ and also the calculated IRM critical density $n_\mathrm{th}$.  This implies that, although the low-density 2D system would invariably undergo a classical percolation transition from a metal to an insulating disordered nonmetal due to the failure of screening in random charged impurity environment, the percolation transition is effectively preempted by the strong Anderson localization crossover defined by the IRM criterion where the coherent quasiparticle transport is no longer possible, thus leading to an effective Anderson insulator \cite{dassarmaTwodimensionalMetalinsulatorTransition2014d}. 
The idea that the 2D MIT is consistent with a quantum Anderson localization crossover defined by the IRM criterion, even in the percolating long-range disorder environment of random Coulomb disorder, is also directly supported by the values of the approximate experimental `critical' crossover resistivity at the transition, $\rho_\mathrm{ex}$, in Table \ref{table:2}, which is always typically of O($h/e^2$) $\sim 26$ k-ohms (within a factor of 2) in Table \ref{table:2}, except for the two samples where the transport measurements are carried out at very high temperatures (and eventual low-$T$ measurements are likely to change the corresponding $\rho_\mathrm{ex}$ closer to 26 k-ohms).  This is, of course, consistent with the IRM criterion where $\rho_\mathrm{th} = \rho_\mathrm{IRM} \sim$ 26 k-ohms. For a WC transition, it is a challenge to explain an experimental $\rho_\mathrm{ex} \sim$ 26 k-ohms, which appears generically almost in all 2D MIT measurements.

(5) The fact that $r_i$ is of O(1), and $r_c$ is typically $<36$ in Table \ref{table:2} again support the AI interpretation of 2D MIT phenomenon.

(6) Finally, the four distinct measures of the disorder strength in Table \ref{table:2}, $\mu$, $n_\mathrm{per}$, $n_\mathrm{i1}$, $n_3$, all tend to indicate that by far the cleanest system studied in the 2D MIT experimental literature is the n-GaAs 94 \cite{sarmaTwoDimensionalMetalInsulatorTransition2005} sample, which, however, manifests 2D MIT at a rather small effective $r_s$ value of 13, far below the QMC prediction of $r_c \sim 36$, thus questioning the validity of equating 2D MIT in ‘clean’ systems necessarily as a WC transition since the cleanest system seems to be undergoing an IRM-controlled AI transition.  In Table \ref{table:3}, we provide an ordered list of all 11 samples showing their ‘cleanliness’ ranking from the top to the bottom, which shows that the 2D MIT is not likely a WC transition since the two cleanest samples \cite{sarmaTwoDimensionalMetalInsulatorTransition2005, manfraTransportPercolationLowDensity2007}
do not manifest the WC transition at all, instead manifesting the AI transition.

\begin{table*}[t]
	\caption{Shows an ordered list of all the samples from the `cleanest'(top) to the `dirtiest'(bottom) along with the MIT crossover densities.}
	\label{table:3}
	\includegraphics[width=\linewidth]{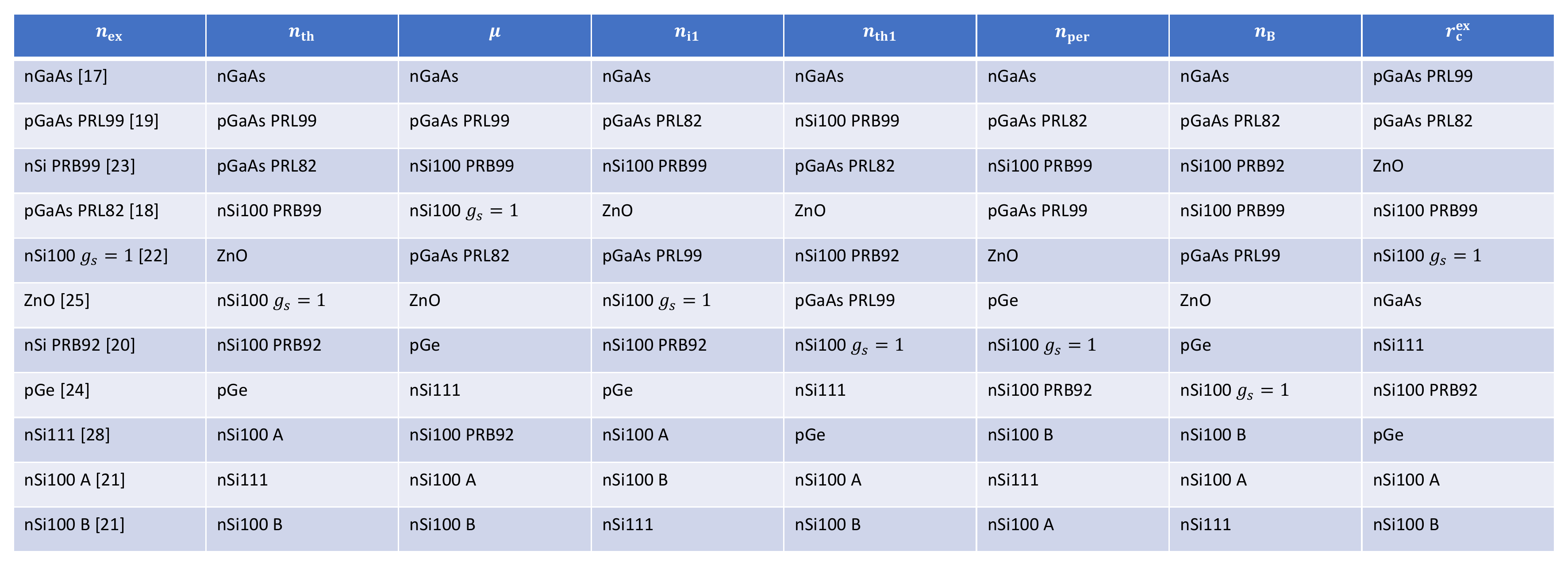}
\end{table*}

\begin{figure}[!htb]
    \centering
    \includegraphics[width=\linewidth]{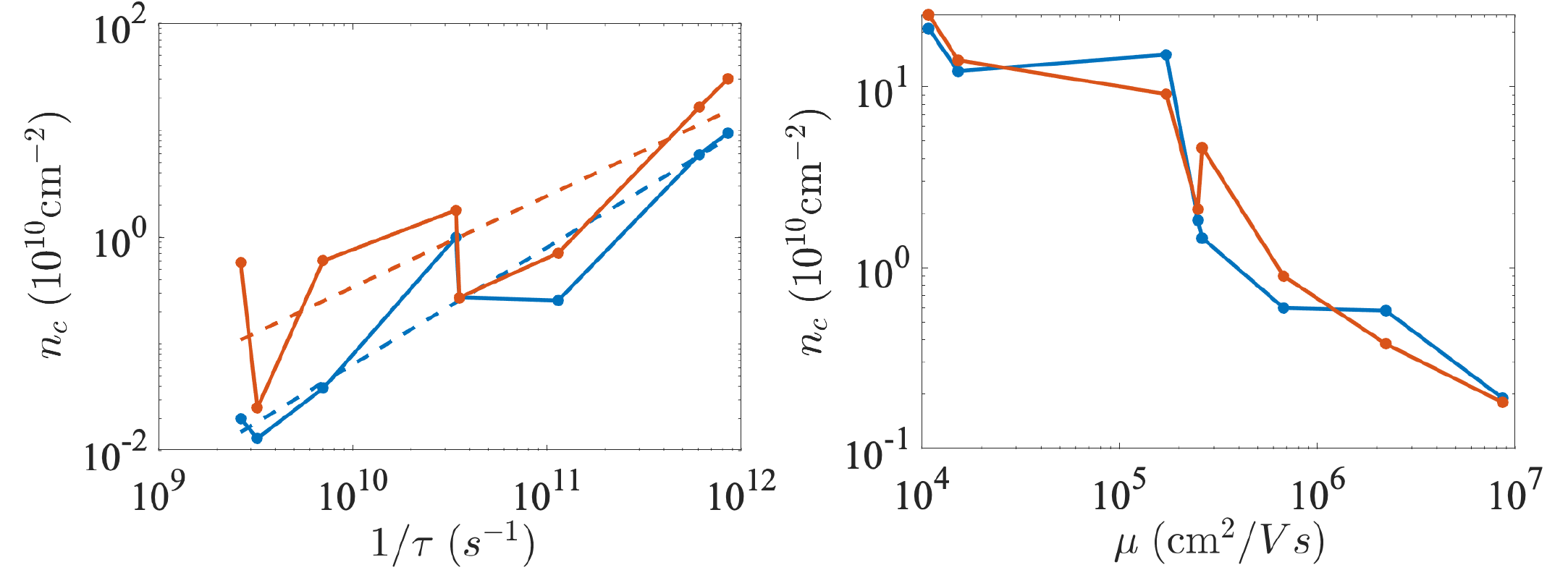}
    \caption{ Shows the experimental (red) and theoretical (blue) plots for the critical 2D MIT density as a function of scattering rate (left panel) or mobility (right panel) deep in the metallic phase.  The dashed straight lines are best fits with the exponent p=.71 (experiment); .67 (theory). Note that the critical 2D MIT density decreases monotonically with decreasing (increasing) scattering rate (mobility) with nothing special happening around the theoretical WC critical density.}  
    \label{fig:5}
\end{figure}

Table \ref{table:3} reinforces our contention that the experimental claims of the WC manifestations in 2D samples arising from the sample ‘cleanliness’ (i.e. relatively small amount of disorder leading to ultra-high mobility) does not hold up to a factual scrutiny.  By far the cleanest two samples, nGaAs \cite{sarmaTwoDimensionalMetalInsulatorTransition2005} and pGaAs 99 \cite{manfraTransportPercolationLowDensity2007} in our list (Table \ref{table:3}), manifest disorder-driven 2D MIT leading to the AI phase and Wigner crystallization. What Table \ref{table:3} suggests is that the experimental critical density for 2D MIT varies primarily with the sample disorder, becoming smaller with decreasing sample disorder, and it is possible for the observed critical density, for some clean samples, purely fortuitously to be low enough to be consistent with the corresponding QMC-predicted 2D WC transition density.  
To make this disorder dependence explicit, we show in Fig.~\ref{fig:5} both the experimental and the theoretical critical density, for the 11 samples studied in the current work, plotted as a function of the (high-density) disorder scattering rate and the mobility (deep in the metallic state).  The abscissa in Fig.~\ref{fig:5} is simply an approximate quantitative measure of the strength of the quenched Coulomb disorder in the system.  In spite of some (expected) fluctuations in the plots, the average qualitative trend is clear: The critical 2D MIT density decreases monotonically with decreasing (increasing) scattering rate (mobility) with nothing special happening around the theoretical WC critical density.  In fact, the 2D MIT critical density varies as a function of the scattering rate according to an approximate power law
\begin{equation} \label{eq:14}
    n_c\sim \left(\frac{1}{\tau}\right)^p
\end{equation}
where $p \sim 0.7$ over a broad range  ($\sim 3$ orders of magnitude) of the scattering rate.  Such a dependence of the critical density on the sample mobility in the context of 2D MIT was first noted empirically in a little-known paper a long time ago \cite{sarachikm.p.DisorderdependenceCriticalDensity2002},
and is consistent with the Ioffe-Regel-Mott criterion under the strong screening approximation at high density \cite{dassarmaTwodimensionalMetalinsulatorTransition2014d, dassarmaScreeningTransport2D2015}. 

We point out that the relationship between mobility $\mu$ and scattering time $\tau$ is the following identity:
\begin{equation} \label{eq:15}
    \mu = \frac{e\tau}{m}
\end{equation}
and the effective mass $m$ varies quite a bit (from 0.07 for nGaAs to 0.4 for pGaAs) among the 11 samples considered in the current work.

\begin{figure}[!htb]
    \centering
    \includegraphics[width=\linewidth]{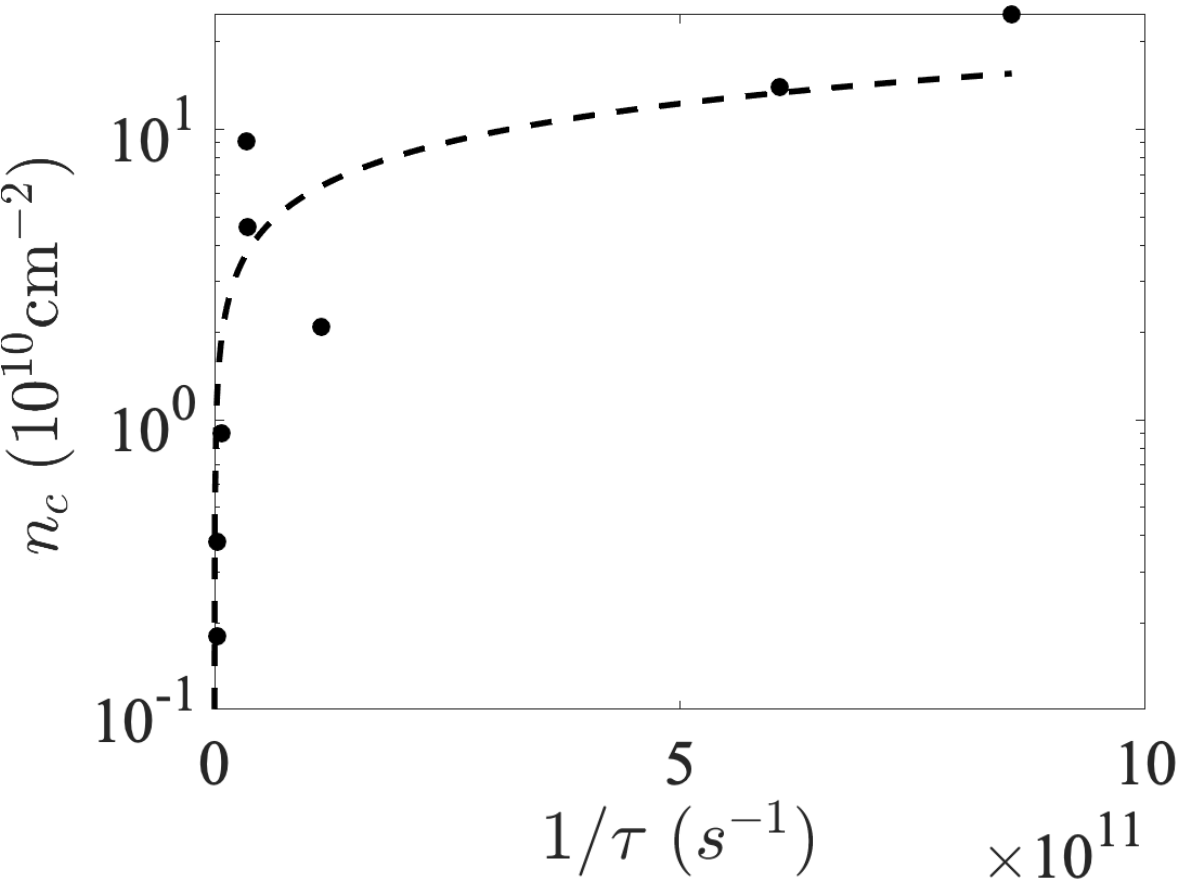}
    \caption{ Shows the extrapolation of the experimental critical density to vanishing disorder with the extrapolated $n_c (1/\tau =0) \sim 10^6 \mathrm{cm}^{-2}$, which implies that the 2D MIT is an AI crossover from an effective metal to a strongly localized Anderson insulator.}  
    \label{fig:6}
\end{figure}

A reasonable empirical question in this context is about the fate of 2D MIT in the limit of vanishing disorder:  Does 2D MIT survive zero disorder?  We can only address this question by extrapolating from the current experimental results (as presented in Table \ref{table:2} or Fig.~\ref{fig:5}, for example), and whether such an extrapolation to zero disorder is meaningful or not is beyond the scope of the current work. In Fig.~\ref{fig:6}, we show a recursive least squares best fit extrapolation of the 2D MIT critical density to vanishing disorder (i.e., $1/\tau$ going to zero, where $\tau$ is the high-density scattering time as measured deep in the metallic phase), finding that the extrapolated $n_\mathrm{c}$ also vanishes (within the numerical accuracy) in the limit of a divergent scattering time!  We find from our numerical extrapolation of the existing 2D MIT experimental data,
\begin{equation}
    n_c=A\left(\frac{1}{\tau}\right)^p+B    
\end{equation}
that $B \sim 10^6 \mathrm{cm}^{-2}$, strongly suggesting (based only on the extrapolation of the experimental 2D MIT data) that $n_c$ indeed vanished in the limit of zero disorder. This is of course completely consistent with the 2D MIT being an AI crossover from an effective metal to a strongly localized Anderson insulator as we claim in the current work.

We mention that once weak localization, which we ignore in our work because it is unobservably small in the experimental temperatures and densities in 2D semiconductor systems, is included in the theory,  the 2D system is strictly speaking always an insulator for all densities in the presence of any finite disorder, but with essentially exponentially long localization length, which therefore has no experimental or physical significance except perhaps at extremely low temperatures. [ PRB90, 125410 (2014)]  Strictly speaking, the 2D MIT crossover being discussed in this paper is from a weakly localized metal for $n<n_c$ to a strongly localized insulator for $n>n_c$.

\section{Conclusion} \label{sec:5}
We present our detailed RPA-Boltzmann theory based analysis of low-temperature density-dependent resistivity in a large number of representative 2D semiconductor layers, finding that the extensively observed density-tuned effective 2D metal-insulator transition is in all likelihood a universal disorder-driven crossover from a high-density effective 2D metal to a low-density strongly localized Anderson insulator. In particular, we find that the claimed interaction-driven transition from a metal to a Wigner crystal based on the very low crossover density for the 2D MIT in some ultraclean samples is inconsistent with the quantitative details of the MIT, and is more likely a transition to the Anderson insulator phase in low-disorder samples which automatically leads to very low transition density.  Our calculated critical density, assuming the MIT to be a disorder-driven crossover to a strongly localized Anderson insulator phase  characterized by the Ioffe-Regel-Mott criterion, agrees with the experimentally reported transition density in essentially all the samples (11 overall), including the samples where the Wigner crystallization has been claimed. In addition, the observed `critical' resistance at the 2D MIT for all 11 samples is approximately consistent with the Ioffe-Regel-Mott criterion, and hence an Anderson localization transition, again ruling out the dominance of Wigner crystallization in any sample, including the ones manifesting MIT at very low carrier densities. We emphasize that our work is not a theory for the localized Wigner glass like low-density localized phase, we only focus on the effective metal-to-insulator crossover showing that the crossover itself is consistent with a disorder induced Anderson localization transition and not a transition to a pristine Wigner crystal as often claimed in the experimental literature.  Our current theory is also not a theory for the effective high-density metallic phase which is essentially a disordered Fermi liquid as has been already studied extensively in the literature.

Our work points to the following appealing physical picture underlying 2D MIT.  The dominant disorder in 2D semiconductor layers is Coulomb disorder, arising from quenched random charged impurities in the environment (our theory includes only Coulomb disorder). At high carrier density, the disorder is strongly screened, leading to weak carrier scattering of well-defined quasiparticles and consequently, effective metallic transport. With decreasing carrier density and associated weakened screening, the effective disorder becomes stronger although the bare disorder in a sample is by definition fixed (e.g. the spatial distribution of random charged impurities) and is independent of the carrier density. This effective carrier density dependent screened Coulomb disorder eventually becomes strong enough at low enough carrier density to drive a higher-density effective metal to an effective strongly localized lower-density Anderson insulator, which is sharp enough to look like a transition (although the MIT is in actuality a crossover from a weakly localized effective metal to a strongly localized Anderson insulator).  The point, which has not been emphasized before in the literature, is that in a given sample, no matter how low the charged impurity density might be (i.e. no matter how ultra-clean the sample might be), the decreasing carrier density will eventually lead to strong effective disorder when the carrier density goes below the impurity density.  Thus, the eventual low-density phase of the 2D semiconductor is always a strongly localized Anderson insulator!  Of course, with decreasing carrier density, the dimensionless interaction coupling $r_s \sim n^{-1/2}$ increases, but the effective Coulomb disorder increases as $n_i/n \sim n^{-1}$, where $n_i$ is the effective charged impurity density.  Thus, eventually, when $r_s$ is large enough at low enough carrier density, Coulomb disorder must prevail over electron-electron interaction.  The QMC estimate for the pristine 2D WC transition at $r_s \sim 36$ is simply irrelevant, unless the system is so clean that the condition $n\gg n_i$ applies at the WC transition point.  This condition is NEVER satisfied in any experimental sample as is obvious from Table \ref{table:2}.  We therefore contend that all reported 2D MIT are Anderson localization crossovers from a high-density effective metal to a low-density strongly localized insulator.

At low densities, correlation effects are obviously strong (along with disorder effects), and our theory becomes progressively poor quantitatively, but the theory remains internally consistent as long as quasiparticles are well-defined.  Thus, the theory, while not being quantitatively very accurate, remains applicable all the way to the MIT (defined by the IRM criterion in our theory) from the metallic side.  One should think of the low-density AI phase (for $n<n_\mathrm{IRM}$) as a strongly correlated and strongly localized insulator, which is a quantum electron glass.  The WC correlations may very well exist in this system for $n<n_\mathrm{WC}$, but such correlations are necessarily short-ranged, and the system is closer to an AI than a WC \cite{vuThermalMeltingQuantum2022}. 
This physical `glassy' picture of low-density disorder-dominated short-range correlated crystallites may be akin to the microemulsion phase hypothesized a long time ago \cite{SpivakIntermediatePhases2003},
and the nonperturbative importance of both disorder and interaction makes this large $r_s$ and strong disorder ($n \sim n_\mathrm{i}$) situation theoretically intractable.
The great advantage of our RPA-Boltmann transport theory is that it makes quantitative predictions for realistic systems taking into account the key physics of density-dependent screened disorder as the underlying mechanism for 2D MIT, which simply cannot be and has not been done for any other theoretical models all of which must assume $n_c$ to be a given experimental parameter.  The reasonable agreement between our theory and experiment for many different 2D systems, as described in great depth in the current work, argue in favor of the approximate validity of our theory.
Doing a reliable and well-controlled theory for such a disordered correlated localized insulator is essentially impossible, and our theory provides a reasonable methodology for extracting the critical density for 2D MIT.  We emphasize that the way to confirm the WC existence is not through transport measurements (which is incapable of distinguishing between AI and WC anyway), but by decisively establishing a property which exists only in a solid/crystal, and not in a liquid.  This could, for example, be the direct observation of the WC lattice structure (or at least, a characteristic wave vector corresponding to the lattice period) and/or observing the phonon modes of the WC.  Establishing these characteristic WC properties necessitate going beyond transport measurements.
There are claims of WC observations in 2D transition metal dichalcogenides, where the physics is very different, and our theory does not apply there \cite{smolenskiSignaturesWignerCrystal2021, zhouBilayerWignerCrystals2021} although the possibility of disorder-induced AI transition in dichalcogenides has also been considered in the literature \cite{ahnDisorderinducedTwodimensionalMetalinsulator2022, ahnTemperaturedependentResistivityDoped2022}. 

Finally, we note that strictly speaking 2D metals do not exist by virtue of the scaling theory of localization \cite{abrahamsScalingTheoryLocalization1979a}. 
This is, however, irrelevant for the 2D MIT considerations since the metal here is an effective metal as appearing experimentally.  At $T=0$ in the thermodynamic limit, the 2D system is always a strongly localized AI even in the presence of infinitesimal disorder.  But our interest is a finite (but, low)-T effective metal, crossing over from a weakly localized insulator to a strongly localized Anderson insulator as the carrier density (effective disorder) decreases (increases).  This crossover is very sharp for Coulomb disorder because of the rapid failure of screening at $n \sim n_\mathrm{IRM}$.  Just as the nonexistence of a true quantum critical 2D localization is irrelevant for the 2D MIT crossover phenomenology, the fact that there cannot be any 2D WC at any finite $T$ is also irrelevant since all that matters for an effective WC is the existence of some spatial order. The physics discussed in this work is strictly speaking crossover physics as observed experimentally and not quantum critical physics which does not exist here  strictly theoretically.

Note added:  A recent experiment \cite{hossainAnisotropicTwodimensionalDisordered2022} claiming Wigner crystallization in 2D nAlAs appeared very recently during the preparation of this manuscript, and, although we believe our conclusion applies to this work, we cannot be definitive since we have not analyzed this system using our transport theory.

\section{Acknowledgement} \label{sec:acknowledgement}
This work is supported by the Laboratory for Physical Sciences.
\clearpage

%


\end{document}